\documentclass[10pt,twocolumn,letterpaper]{article}

\usepackage[T1]{fontenc}
\usepackage[utf8]{inputenc}
\usepackage{amsmath}
\usepackage{newtxtext,newtxmath}
\usepackage[margin=0.75in,columnsep=0.25in]{geometry}
\usepackage{graphicx}
\usepackage{booktabs}
\usepackage[round,authoryear]{natbib}
\usepackage{algorithm}
\usepackage{algpseudocode}
\usepackage{xcolor}
\usepackage[switch,mathlines]{lineno}
\usepackage{authblk}
\usepackage[colorlinks=true,allcolors=blue!60!black]{hyperref}

\newcommand{\ind}{\mathbf{1}}
\newcommand{\E}{\mathbb{E}}
\newcommand{\Prb}{\mathbb{P}}
\newcommand{\Sset}{\mathcal{S}}

\title{The Conditional Value of Costly Information in Collective-Risk Dilemmas}

\author[1,2,3*]{Elias Fern\'andez Domingos}
\author[4]{Genki Ichinose}
\author[5]{Wakaba Tateishi}
\author[1,2,3,6]{Tom Lenaerts}
\affil[1]{AI Lab, Vrije Universiteit Brussel, Brussels, Belgium}
\affil[2]{MLG, Universit\'e Libre de Bruxelles, Brussels, Belgium}
\affil[3]{FARI Institute,  Universit\'e Libre de Bruxelles - Vrije Universiteit Brussel, Brussels, Belgium}
\affil[4]{Shizuoka University, Hamamatsu, Japan}
\affil[5]{Hokkaido Musashi Women's University, Hokkaido, Japan}
\affil[6]{Center for Human-Compatible AI, UC Berkeley, USA}
\affil[*]{Corresponding author: \texttt{elias.fernandez.domingos@vub.be}}

\date{Preprint}

\begin{document}

\twocolumn[
  \begin{@twocolumnfalse}
    \maketitle
    \begin{abstract}
Many high-stakes collective problems, e.g., climate mitigation and pandemic response, share a structure in which risk materializes unless a group threshold is met. This threshold is often uncertain, but uncertainty can be reduced at a cost. We study this information cost tradeoff in an evolutionary game theoretic model of social learning in the collective-risk dilemma, where individuals must cooperate to reach a possibly uncertain threshold to prevent collective loss. Individuals may pay a fixed cost to observe the threshold's true value and condition their cooperation on it. Across four threshold priors, we characterize risk levels at which cooperative equilibria emerge and show that low cost information increases group success. However, this effect is reversed at higher costs. We define a private \emph{value of information} and show that information is worth buying only when it is considerably cheaper than contributing. When purchased information is shared with the group, the cooperative tipping point shifts to lower risk levels. Despite bearing a private cost, above this tipping point buyers earn more in expectation than information free-riders. We then study these dynamics when information diffuses through a network, allowing nodes to relay it. Although fewer individuals pay for information, relaying increases group success, and noisy relaying does not reduce it. By contrast, false information does, especially when it originates at the hubs of heterogeneous networks, and large attacks can bring collective action close to collapse. Buyers act as a firewall against disinformation, confining it to source neighbourhoods, and once disinformation sources are removed, the population recovers. The value of costly information thus depends on its price, the level of risk, and how it is shared, and, in our setting, the effort of acquiring it also protects the population against disinformation.

    \end{abstract}
    \vspace{1.5em}
  \end{@twocolumnfalse}
]

\section{Introduction}

Climate action, pandemic response, the exploitation of finite natural resources, and the spread of disruptive technologies such as AI are prominent examples of collective-risk social dilemmas. Steering cooperation in such dilemmas is a core problem for AI and
multi-agent systems research \citep{RolnickEtAl2022, ShiWangFang2020}.

These situations may be operationalized in game theory through the collective-risk dilemma (CRD) \citep{Milinski2008}. Here, we adopt a one-shot, evolutionary-game-theoretic (EGT) version of this dilemma \citep{SantosPacheco2011, VasconcelosSantosPacheco2013,
CoutoPachecoSantos2020}, in which a group of individuals must decide whether to cooperate by contributing to a public good or to defect and withhold their contribution. If the group reaches a collective threshold number of cooperators, all participants keep their remaining endowment; otherwise, all participants lose it with probability equal to the collective risk $r$. Defectors thus avoid the cost of preventing the disaster, but insufficient cooperation leads the group to its demise.

In CRD, individuals need to coordinate around the collective threshold. However, when the threshold is uncertain, this focal point for coordination is lost. We argue here that there may be many sources of such uncertainty. In climate action, for instance, \citet{IPCC2018} documents substantially greater climate-related risks at 2\textdegree{}C of warming than at 1.5\textdegree{}C, while uncertainty about the location of climate tipping points can undermine coordination on emissions reductions \citep{BarrettDannenberg2012}. Such uncertainty is \emph{inherent} to the underlying system and can be reduced only at a cost, for instance through scientific research. Experiments show that cooperation is sensitive to whether threshold uncertainty falls below a critical range \citep{BarrettDannenberg2014}. A second, and arguably more pressing, source of threshold uncertainty is not inherent but manufactured by misinformation. For instance, during the COVID-19 pandemic, nearly three-quarters of U.S. adults reported exposure to conflicting
information about COVID-19 \citep{Nagler2020}, illustrating the conflicting messages that can complicate protective decisions. Unlike inherent uncertainty, which shrinks as science advances, this manufactured variety is actively sustained by actors who benefit from a confused public \citep{Lazer2018}, a problem compounded by language models whose misinformation can be harder to detect than human-written misinformation \citep{ChenShu2024}. Moreover, most individuals do not acquire such information themselves but receive it from others, who relay it through
social networks in which false news spreads farther and faster than true news \citep{VosoughiRoyAral2018}.

As outlined above, this structure recurs well beyond climate action. This
raises our two central questions: first, does purchasing or investments in information about a shared threshold help or harm the group's ability to avoid collective disaster? And second, can individuals who pay such a cost for truthful information, and relay it to others, increase the resilience of the system to misinformation and disinformation? Prior work has studied how \emph{exogenous} forms of uncertainty (e.g., impact, timing, ambiguity) affect cooperation in the CRD \citep{FernandezDomingos2020,PachecoVasconcelosSantos2014,Dannenberg2015}, but whether agents should be given the option to acquire information and relay it has, to our knowledge, not been examined in evolutionary models of the CRD.

Answering these questions in full generality is intractable. We instead study them through a
deliberately simplified EGT model of social learning \citep{SantosPacheco2011,FernandezDomingos2023}. We consider a homogeneous population facing a single uncertain threshold governed by one of four stylized prior distributions, and a minimal set of four strategies that either act unconditionally or pay a fixed cost to observe the threshold
and act on it conditionally. This isolates the causal effect of information cost and threshold uncertainty on group cooperation from confounds (e.g., heterogeneous endowments, repeated interaction, communication) that would otherwise obscure it. We then let information bought by one group member be shared with the whole group, and finally
place the population on a network on which information diffuses from buyers through those who act on it, including imperfect relays, a malicious relay, and external sources of noisy or false information.

Our contributions are: (1) the first EGT model of the one-shot CRD in which agents can pay a fixed cost to resolve threshold uncertainty and act on it conditionally; (2) a characterization, across four threshold prior distributions and the full risk range, of when costly information raises group success and when it backfires, explained by the value of information to an individual, which we show is bounded by the cost of cooperation; (3) the finding that sharing information within groups turns its acquisition into a collective
good that shifts the tipping point to lower risk; (4) an analysis of information diffusion on networks, showing that relaying raises group success even though fewer individuals pay for information, and that noisy relaying does not harm collective success whereas false information does, especially when it originates at hubs. We also show that individuals who pay for information act as a firewall, confining disinformation to the neighbourhood of the sources that generate it, and that the population is able to recover once these sources are removed.

\section{Related Work}

Collective-risk dilemmas (CRDs) have been thoroughly studied both experimentally \citep{Milinski2008, TavoniEtAl2011} and theoretically \citep{SantosPacheco2011, PachecoVasconcelosSantos2014, VasconcelosSantosPacheco2013, CoutoPachecoSantos2020}. The one-shot CRD introduced by \citet{SantosPacheco2011} has since been extended to model bottom-up, self-organized institutions \citep{VasconcelosSantosPacheco2013,garrido2025new,pacheco2026revenue} and graduated punishment \citep{CoutoPachecoSantos2020}. The literature has also examined how different sources of uncertainty affect cooperation. Impact and timing uncertainty help in certain cases \citep{FernandezDomingos2020,  PachecoVasconcelosSantos2014}, while threshold uncertainty \citep{BarrettDannenberg2012, PachecoVasconcelosSantos2014} generally hurts it, and its negative effect is particularly severe once the threshold's own distribution becomes ambiguous rather than merely unknown \citep{Dannenberg2015}. Notably, \citet{BarrettDannenberg2014} show that restoring cooperation requires narrowing this uncertainty to a small range around the true threshold, precisely what our costly-information mechanism lets agents do endogenously, at the individual rather than the institutional level.

A separate line of work asks what happens when the uncertainty is not merely unresolved but actively distorted. For example, misinformation about co-players' actions has been shown to erode cooperation in evolutionary models outside the CRD \citep{Meng2023}, where false reports that defectors cooperate distort expected payoffs and migration decisions. Here, we test how noise and deliberate misinformation interact with a costly-information mechanism in the CRD.

The possibility of paying a cost to transform uncertainty into risk, or to eliminate it entirely, has not been previously studied in theoretical models of the CRD. Outside the CRD, endogenous and costly information acquisition has been studied game-theoretically in coordination games more broadly
\citep{MyattWallace2012}, building on \citet{MorrisShin2002}, who found that the social value of (more) public information in coordination games is not always positive. \citet{VanDriesscheEtAl2024} and \citet{XuStangenberg2025} are the closest references for our work. \citet{VanDriesscheEtAl2024} let participants in a public goods game pay to convert an unknown probability distribution into a known one. There, participants are willing to pay for information in more than half of the decisions, and the authors find that willingness to pay is predicted by individual ambiguity attitude rather than a uniform response. \citet{XuStangenberg2025} let participants in a threshold public goods game pay to acquire a private, exact signal about the uncertain threshold, finding provision rises when acquisition is available and rises further when the acquisition decision itself is made observable to other players.

Population structure is known to change the evolution of cooperation \citep{OhtsukiEtAl2006}, including in public goods games where each individual plays in the groups centred on itself and on its neighbours
\citep{SantosSantosPacheco2008}. In the CRD, heterogeneous networks enlarge the cooperative equilibrium when fitness accumulates over all the games an individual plays \citep{SantosPacheco2011}. At the same time, behaviour and information also spread through social networks \citep{Centola2010}, and false news spreads farther and faster than true news \citep{VosoughiRoyAral2018}. The literature identifies here that the key responses include verifying and correcting information and flagging misleading content \citep{TreenWilliamsONeill2020}. To our knowledge, the interplay between the diffusion of costly threshold information and cooperation in a CRD, as we model here, has not been studied.

Beyond the evolutionary-game-theoretic framework adopted here, multi-agent reinforcement learning has been used to study CRDs with diversity in individual risk \citep{Merhej2022} and broader collective-action problems such as climate negotiations among nations
\citep{ZhangEtAl2025}.

\section{Model}

\subsection{The collective-risk dilemma with costly information}

We formalize the cost-uncertainty trade-off discussed in the Introduction with an evolutionary-game-theoretic model where a population plays a one-shot collective-risk dilemma (CRD) against an uncertain threshold, and individuals may choose whether to pay to observe it before acting.

\paragraph{Setup.}
We consider a population of $Z$ individuals randomly assigned to groups of size $N$ to play a one-shot CRD. Each individual holds an endowment $b>0$. A cooperator contributes the fraction $c\in[0,1]$ of
it (paying $cb$), a defector contributes nothing. Let $M\in\{0,\ldots,N\}$ denote the minimum number of cooperators required for the group to succeed. If the group has at least $M$ cooperators it succeeds and every player keeps the endowment $b$; otherwise each player loses it with probability $r\in[0,1]$. Since contributions are realized regardless of the outcome, cooperators pay a cost even when the group fails.

The threshold $M$ is drawn independently in each interaction from one of four possible discrete priors
$p(M)$: \emph{uniform}, \emph{gaussian}, \emph{low-threshold}, and \emph{high-threshold} (Supporting Information (SI), Figure~\ref{fig:si-discrete-threshold-priors}). The uniform prior represents complete ignorance about the threshold's location, the gaussian prior concentrates probability mass around the
midpoint, and the low- and high-threshold priors place most of their mass on low and high thresholds, respectively.

Additionally, here, an individual may pay an information cost $c_Ib\geq0$ to observe the realized threshold.

\paragraph{Strategies.}
We adopt a set of four strategies, $\Sset=\{\mathrm{UD},\mathrm{UC},\mathrm{CH},\mathrm{CL}\}$. The two unconditional strategies ($\mathrm{UD}$, $\mathrm{UC}$) do not purchase information. The two conditional strategies ($\mathrm{CH}$, $\mathrm{CL}$) pay $c_Ib$, observe the realized threshold, and condition their action on its relation to a fixed reference threshold
$M_f$. Unless stated otherwise, we fix the reference at $M_f=4$, which is also the midpoint of the threshold support, for all priors.

We define the cooperation indicator of strategy $i\in\Sset$ by
\[
q_i(M)=
\begin{cases}
0, & i=\mathrm{UD},\\
1, & i=\mathrm{UC},\\
\ind\{M\geq M_f\}, & i=\mathrm{CH},\\
\ind\{M<M_f\}, & i=\mathrm{CL}.
\end{cases}
\]

The conditional strategies (conditionally high -- $\mathrm{CH}$ -- and low -- $\mathrm{CL}$) act here as simplified proxies of how a decision-maker that has an internal aspiration about what the collective target should be ($M_f$) compares it to the information it receives and acts on it.

\paragraph{Payoffs.}

For a group with $k\in\{0,\ldots,N\}$ cooperators and threshold $M$, the expected payoff of a defector and a cooperator are
\begin{align}
\pi_D(k,M)&=b\left[(1-r)+r\,\ind\{k\geq M\}\right],\\
\pi_C(k,M)&=\pi_D(k,M)-cb.
\end{align}

Conditional strategies additionally pay the information cost $c_Ib$, captured by the indicator $I_i=\ind\{i\in\{\mathrm{CH},\mathrm{CL}\}\}$.

Suppose a focal player using strategy $i$ faces $j$ cooperating co-players among its $N-1$ group-mates, so the total number of group cooperators is $k=j+q_i(M)$. Then, the focal player's expected payoff, conditional on $j$ and $M$, is:
\begin{equation}
\label{eq:primitive-payoff}
\pi_i(j,M)
=
\pi_D\bigl(j+q_i(M),M\bigr)
-q_i(M)cb-I_ic_Ib.
\end{equation}

\paragraph{Notation convention.}
From here on, we suppress state arguments when clear from context. For example, in the definitions below $\Pi_i$ denotes either $\Pi_i(\mathbf{x})$ (infinite population) or $\Pi_i(\mathbf{K})$ (finite population), likewise for $\eta$ and $\overline\Pi$.

\subsection{Evolutionary dynamics}
\label{sec:evolutionary-dynamics}

\paragraph{Infinite population.}
Let $x_i$ be the frequency of strategy $i$, with
\[
x_i\geq0,\qquad \sum_{i\in\Sset}x_i=1.
\]
The deterministic dynamics are given by the replicator equation \citep{HofbauerSigmund1998}
\begin{equation}
\dot x_i=x_i\left(\Pi_i-\overline\Pi\right),
\qquad
\overline\Pi=\sum_{\ell\in\Sset}x_\ell\Pi_\ell.
\end{equation}

For a realized threshold $M$, each of the $N-1$ co-players cooperates independently with probability
\[
\overline q(M)=\sum_{\ell\in\Sset}x_\ell q_\ell(M),
\]
so the number of cooperators is Binomially distributed with this parameter. Thus, the expected payoff of strategy $i$ is given by:
\begin{equation}
\label{eq:infinite-payoff}
\begin{aligned}
\Pi_i
={}&
\sum_{M=0}^{N}p(M)
\sum_{j=0}^{N-1}
\binom{N-1}{j}
\overline q(M)^j \\
&\times\bigl(1-\overline q(M)\bigr)^{N-1-j}
\pi_i(j,M).
\end{aligned}
\end{equation}

\paragraph{Two-strategy reduction.}
Restricting to the two unconditional strategies $\{\mathrm{UD},\mathrm{UC}\}$ recovers the original model of \citet{SantosPacheco2011}, used for the mutation-free replicator analysis in Figure~\ref{fig:combined-optima}.

\paragraph{Finite population (social learning).}
Let $\mathbf{K}=(K_{\rm UD},K_{\rm UC},K_{\rm CH},K_{\rm CL})$ be the vector of strategy counts, with $\sum_{i\in\Sset}K_i=Z$. Since the population is finite, the $N-1$ co-players of a focal individual are
instead sampled \emph{without} replacement from the remaining $Z-1$ individuals, so the number of cooperators follows a Hypergeometric distribution, and the resulting expected payoff $\Pi_i(\mathbf{K})$ takes the same form as Equation~\eqref{eq:infinite-payoff} with this Hypergeometric distribution in place of the Binomial (SI, Equation~\eqref{eq:finite-payoff}).

At each update, a focal individual $A$ and a role model $B$ are sampled. The focal individual adopts the role model's strategy with Fermi probability $p_{A,B}=(1+\exp\left[-\beta\left(\Pi_B-\Pi_A\right)\right])^{-1}$ \citep{TraulsenNowakPacheco2006}, where $\beta\geq0$ is the intensity of selection.

With mutation or exploration probability $\mu$, the updating individual instead selects uniformly among the alternative strategies, $\Sset\setminus\{s_A\}$, where $s_A$ is its current strategy. This produces an ergodic Markov chain when $\mu>0$, with stationary
distribution $\rho(\mathbf{K})$.

\paragraph{No-information (NI) baseline.}
Throughout the Results, in the finite-population NI baseline, we consider four strategies, where UC and UD are duplicated, instead of the previous 2-strategy model. We do this because mutation acts at the level of strategies, which makes a comparison between a 2 and a 4 strategy model confounded. By creating the duplicates, we ensure that all do is modify the strategies present in the system, not their number, so that any effect is produced by the new strategies and not just because the system has more strategies (SI, Section~\ref{sec:si-ni}).

\paragraph{Population-level dependent variables.}
Beyond the payoffs that drive the dynamics, we further report three population-level quantities, using the notation convention above to cover both frameworks (SI, Section~\ref{sec:si-variables}, gives their exact expressions). Group success $\eta$ is the probability that a randomly drawn group of $N$ members meets the threshold, evaluated at the equilibrium $\mathbf{x}^*$ (replicator) or as the finite-population stationary average. Cooperation frequency $f_{\rm coop}$ and information-purchase frequency $f_{\rm info}$ follow directly from the strategy frequencies and each conditional strategy's chance of cooperating under the prior (with $M_f=4$). The population-average expected payoff
$\overline\Pi=\sum_{i\in\Sset}x_i\Pi_i$ decomposes as $\overline\Pi=(1-r)b+rb\,\eta-cb\,f_{\rm coop}-c_Ib\,f_{\rm info}$, that is, the endowment kept even when groups fail, plus the endowment saved by group success, minus the costs of cooperation and of information, with $\eta$, $f_{\rm coop}$, $f_{\rm info}$, and $\overline\Pi$ denoting stationary averages in the finite-population case.

\paragraph{Value of information.}
We define here the value of information as the private value of observing the threshold. Consider a focal individual in population state $\mathbf K$, and let $\mathrm{Piv}(M)$ be the probability that exactly $M-1$ of its $N-1$ co-players contribute at threshold $M$, so that its own contribution decides
the outcome. From Equation~\eqref{eq:primitive-payoff}, contributing instead of defecting changes the focal individual's expected payoff at threshold $M$ by
\begin{equation}
\label{eq:marginal-contribution}
\Delta\Pi(M)=b\bigl[r\,\mathrm{Piv}(M)-c\bigr].
\end{equation}
(derivation in SI, Section~\ref{sec:si-voi-derivation}). Since an uninformed individual must take the same action at every threshold, its maximum gain over defection is given by the maximum of always and never contributing ($\max\{\E_p[\Delta\Pi(M)],0\}$). Contrarily, an informed individual can contribute exactly where $\Delta\Pi(M)>0$, gaining $\E_p[\max\{\Delta\Pi(M),0\}]$. The difference is the value of information \citep{Howard1966},
\begin{equation}
\label{eq:voi}
V=\E_p\bigl[\max\{\Delta\Pi(M),0\}\bigr]-\max\bigl\{\E_p[\Delta\Pi(M)],0\bigr\}\geq0,
\end{equation}
that is, the largest acquisition cost an individual would accept, holding the rest of the population fixed. Thus, information is only valuable ($V>0$) if it changes the individual's decision, and $\Delta\Pi(M)$ is positive only at thresholds where $\mathrm{Piv}(M)>c/r$. Moreover, writing $\Delta\Pi=cb\,(\theta-1)$ with the pivotality-to-cost ratio $\theta(M)=r\,\mathrm{Piv}(M)/c$, Equation~\eqref{eq:voi} becomes
\begin{eqnarray}
\label{eq:voi-bound}
V &=& c\,b\,\varphi, \nonumber\\
\varphi &=& \min\bigl\{
\E_p[(\theta-1)^{+}],
\E_p[(1-\theta)^{+}]
\bigr\} \nonumber\\
&\leq& \Prb_p\bigl(\theta(M)<1\bigr) \leq 1.
\end{eqnarray}
where $x^{+}=\max\{x,0\}$. Hence, information pays either by enabling contributions that are worthwhile ($\theta>1$) or by avoiding those that are not ($\theta<1$). In this way, information is never worth more than the cost of cooperation, and acquisition pays when $c_I<\varphi\,c$. In Figures~\ref{fig:baseline-comparison}C and~\ref{fig:shared-info}C we use $V/c_I$, with $V/c_I>1$ when the value of information is larger than its price.

Finally, the value $V$ assumes that an informed individual uses the threshold
optimally, contributing when $\Delta\Pi(M)>0$. However, CH and CL use $M$ through a fixed reference $M_f$, so the value they realize, $V_i=\E_p[q_i(M)\Delta\Pi(M)]-\max\{\E_p[\Delta\Pi(M)],0\}\leq V$, can be negative when they contribute if $\Delta\Pi(M)<0$ or when they do not if $\Delta\Pi(M)>0$.

\subsection{Sharing information within groups}
In the model above, information can be used only by the individual who buys it. To ask whether information acquisition then becomes a collective good, we let information bought by any group member be shared with the whole group. We consider five strategies, $\Sset=\{\mathrm{UD},\mathrm{UC},\mathrm{BH},\mathrm{FH},\mathrm{BD}\}$. UD and UC remain unchanged. A buyer (BH) pays $c_Ib$, shares the threshold with its group, and contributes iff $M\geq M_f$, while an information free-rider (FH) does not buy, and instead, contributes iff $M\geq M_f$ when at least one group member bought the threshold, and defects otherwise. An information provider (BD) buys and shares the threshold but never contributes. BH and FH follow the same rule and differ only in whether they pay, so comparing them measures whether those who acquire information are also essential to group success. BD separates providing information from contributing. As a benchmark, with the same five strategies, we switch sharing off, that is, in this case FH never holds information and acts as UD, and BD pays for information nobody uses. Payoffs follow Equation~\eqref{eq:primitive-payoff} for every group composition, and we solve the stationary distribution over all $\binom{Z+|\Sset|-1}{|\Sset|-1}\approx4.6\times10^6$ population states ($c_I=0.01$, $\beta=3$, $\mu=0.01$). Per-strategy payoffs and pivotal-contribution rates are computed in expectation exactly. We measure the private value of buying as in Equation~\eqref{eq:voi}. With sharing, buying also informs the free-riders in the group, and it has no added value when a co-player already buys.

\subsection{Information diffusion on networks}
We place the population on a random 7-regular network ($Z=100$), so that each node and its neighbours form a group of $N=8$. Similarly to \citet{SantosSantosPacheco2008}, every node plays in the group centred on itself and in those centred on its neighbours, but its payoff is the average instead of the sum over those groups. A single threshold $M$, drawn from $p(M)$, applies to all groups simultaneously. Before players act, information diffuses, so that buyers (BH, BD) learn $M$ and tell all their neighbours, and every informed BH or FH relays what it received to each uninformed neighbour with probability $\lambda$, where $\lambda=0$ restricts information to the buyers' neighbours. Additionally, nodes cannot tell whether the message originates from a buyer, and each node acts on the first message it receives (see SI for the pseudo-code).

In this setup, we consider three conditions: (i) under noisy relaying, each relay adds Gaussian noise of standard deviation $\sigma$ to the threshold (rounded to an integer), so misunderstandings accumulate along the cascade; (ii) under a malicious relay, one non-evolving node that never contributes always tells its neighbours that the threshold is $M=0$, and we place it either at the node of highest betweenness centrality or at random; and (iii) external sources are injected either at the weakest points of the network -- the nodes whose neighbourhood contains the largest share of information free-riders -- or at random nodes. Noisy sources report the true threshold plus Gaussian noise of standard deviation $2$, and biased sources report instead $M=0$. On degree-heterogeneous networks, sources are also placed at hubs. Like malicious relays, sources are fixed nodes that never contribute, so part of their damage would occur even if they sent no message. We therefore compare each attack with the same number of silent sources, fixed nodes that never contribute and send nothing. 

Finally, strategies update asynchronously. At each step, a random node either mutates to another strategy with probability $\mu$ or imitates a random neighbour with the Fermi probability of Section~\ref{sec:evolutionary-dynamics} \citep{TraulsenNowakPacheco2006,SzaboFath2007} ($\beta=3$, $\mu=0.01$, $c_I=0.01$). A node's fitness is computed as the expected average payoff over the games the individual plays in its neighbourhood. We compute the expectation over the threshold distribution and over the possible diffusion cascades (the final state of information for each node in the network). When we make message diffusion stochastic (that is when $0<\lambda<1$, there is noise, or sources compete with buyers), this expectation is estimated from 16 independent cascades. Furthermore, we estimate stationary averages of the dynamics from independent chains started from dispersed initial states on 5 (3 for the relay-noise experiment) network realisations. 

Additionally, to study transients, we run ensembles of replicates through a fixed sequence of phases: we start with a population of UD and UC only, then the information strategies become available, afterwards a malicious relay or external sources are added, and finally they are removed. In this way, we can observe how the different interventions affect the system. We also measure damage and recovery by using an ensemble in which the attack never happens as counterfactual. As in the well-mixed analysis, we run a control with the same five strategies but without information, in which buyers behave exactly as UC and the other two information strategies as UD, to ensure that the observed effects are driven by information and not by the larger strategy set, which changes the mutation process. We repeat the network experiments on degree-heterogeneous Barab\'asi--Albert networks of the same size (SI, Section~\ref{sec:si-network}).

\section{Results}

\begin{figure*}[t!]
  \centering
  \includegraphics[width=0.99\textwidth]{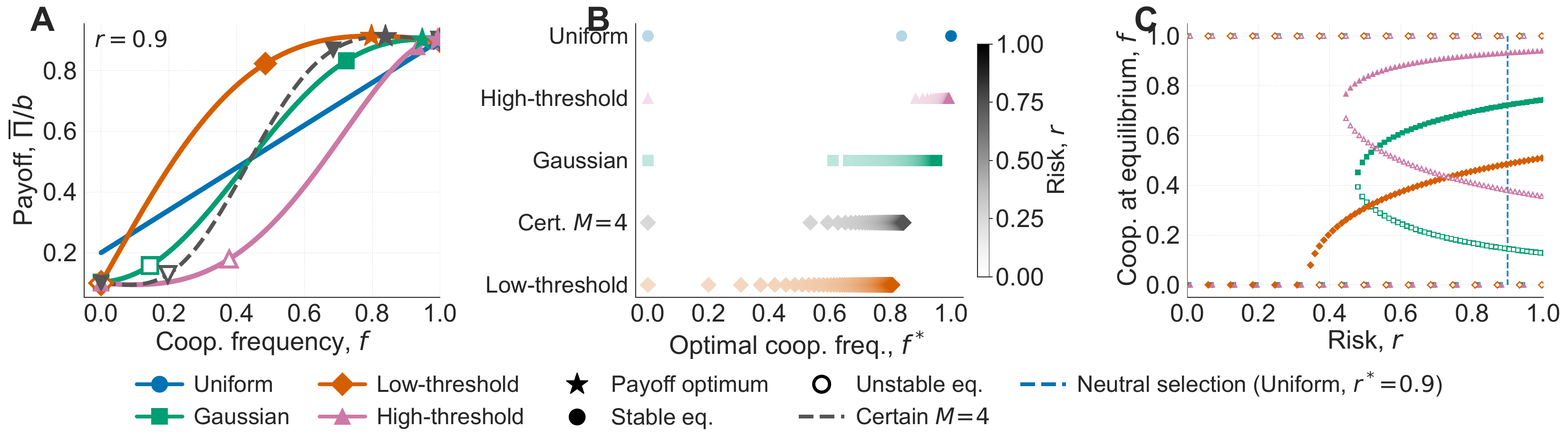}
  \caption{Effect of prior distribution and risk on population equilibria.
  \textbf{(A)} Population-average
  payoff $\overline{\Pi}/b$ as a function of cooperation frequency $f$ at
  $r=0.9$. Filled markers indicate stable equilibria, open
  markers unstable ones, and stars the payoff-maximizing optimum,
  for each prior and the certain-threshold reference ($M=4$).
  \textbf{(B)} Payoff-maximizing cooperation frequency
  $f^*$ as a function of risk, per prior (color encodes $r$, per the colorbar),
  with a certain-threshold reference ($M=4$). \textbf{(C)}
  Bifurcation diagram: equilibria vs.\ $r$ per prior (dashed vertical
  segments mark neutral-selection points).
  ($N=8$, $c=0.1$, $b=1$; $r=0.9$ in panel A)}
  \label{fig:combined-optima}
\end{figure*}

\begin{figure*}[t!]
  \centering
  \includegraphics[width=0.95\textwidth]{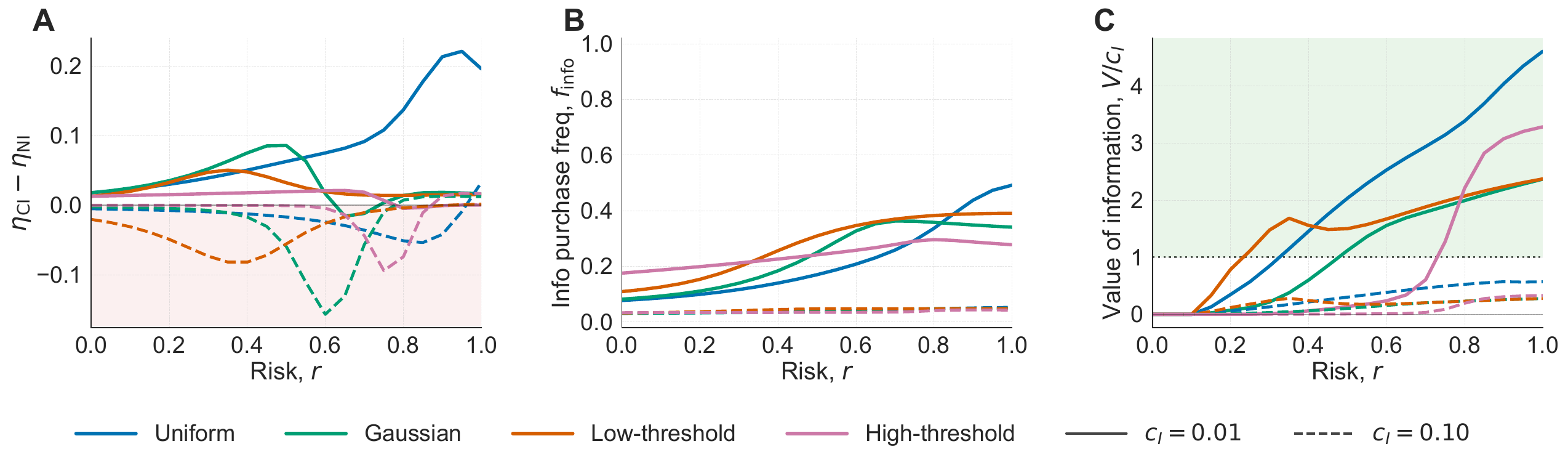}
  \caption{Value of costly information (CI) relative to a
  no-information (NI) baseline, across the four
  threshold priors,
  and two information-cost levels ($c_I=0.01$
  solid, $c_I=0.10$ dashed). \textbf{(A)} Change in group
  success, $\eta_{\rm CI}-\eta_{\rm NI}$, as a function of risk $r$. The red region indicates where CI performs worse than NI.
  \textbf{(B)} Frequency with which CI pays to observe the threshold.
  \textbf{(C)} Value of information relative to its price, $V/c_I$
  (Equation~\ref{eq:voi}), for an individual in the stationary CI population
  evolved at each price. The green region ($V/c_I>1$) indicates when acquisition
  is privately beneficial. Changes in cooperation and payoff are shown in
  Figure~\ref{fig:si-coop-payoff-difference}.
  ($Z=100$, $\beta=3$, $\mu=0.01$, $N=8$, $c=0.1$, $b=1$, $M_f=4$)}
  \label{fig:baseline-comparison}
\end{figure*}

\paragraph{Risk and the prior distribution determine cooperative equilibria.}

Figure~\ref{fig:combined-optima} characterizes the coordination problem before costly information is available. At $r=0.9$, both the payoff-maximizing cooperation frequency and the equilibria of the two-strategy replicator dynamics differ substantially across threshold priors (Figure~\ref{fig:combined-optima}A). Under the low-threshold prior, full defection is unstable and the population settles at an inefficient equilibrium. The gaussian and high-threshold priors are instead bistable. Under the uniform prior, a contribution is pivotal with probability $1/(N+1)$ whatever the cooperation frequency, so the dynamics have a critical risk
$r^*=(N+1)c$, where every $f$ is an equilibrium. Below $r^*$ defection dominates, above it cooperation dominates, and at $r^*$ ($=0.9$ here) selection is neutral.

Across the full risk range, both the payoff-maximizing cooperation frequency $f^*$ and the replicator equilibria remain prior-dependent (Figure~\ref{fig:combined-optima}B--C). Each prior undergoes a transition to cooperation at a different critical risk reflecting how the distribution of probability mass across possible thresholds changes the underlying coordination problem.

\paragraph{The value of costly information depends on price, risk, and the prior.}

Figure~\ref{fig:baseline-comparison} compares the costly-information model (CI) against the
no-information baseline (NI), in which UD and UC are duplicated, in a finite population
with $Z=100$.

When information is cheap ($c_I=0.01$), CI increases group success relative to NI across nearly the entire risk range for every prior (Figure~\ref{fig:baseline-comparison}A). However, while cooperation increases over most of the risk range, payoffs stay close to the baseline for all priors except the uniform one (Figure~\ref{fig:si-coop-payoff-difference}). This reverses when information is
expensive ($c_I=c=0.10$), which follows from the corollary of Equation~\eqref{eq:voi-bound} that the value of information can never be worth more than the cost of cooperation. The small payoff gains of cheap information occur in spite of strategies that acquire information reaching about 30--50\% of the population, depending on the prior, whereas few individuals buy information when it is expensive (Figure~\ref{fig:baseline-comparison}B).

Figure~\ref{fig:baseline-comparison}C shows that such cheap information should still pay off for higher levels of risk and for all priors information is valuable (see the green region), since $V/c_I>1$ once risk is high enough for pivotal contributions to be worth their cost ($r\,\mathrm{Piv}(M)>c$ at some thresholds), which happens first for the low-threshold and last for the high-threshold prior. This result indicates that the information-purchasing heuristics do not take advantage of the full value of information. In fact, CH and CL earn less than the best unconditional action at every risk and prior ($V_i<0$), even before paying for information, because they respond to the position of $M$ relative to a fixed reference rather than to whether their contribution is likely to be pivotal. That is, information is acquired, but used by a rule that ignores whether contributing decides the outcome. Nonetheless, whether contributions are individually pivotal understates the collective role of information, and the contributions made by these informed strategies are jointly decisive for a much larger share of successful groups (SI, Section~\ref{sec:si-collective-pivotality}).

\begin{figure*}[t!]
  \centering
  \includegraphics[width=0.98\textwidth]{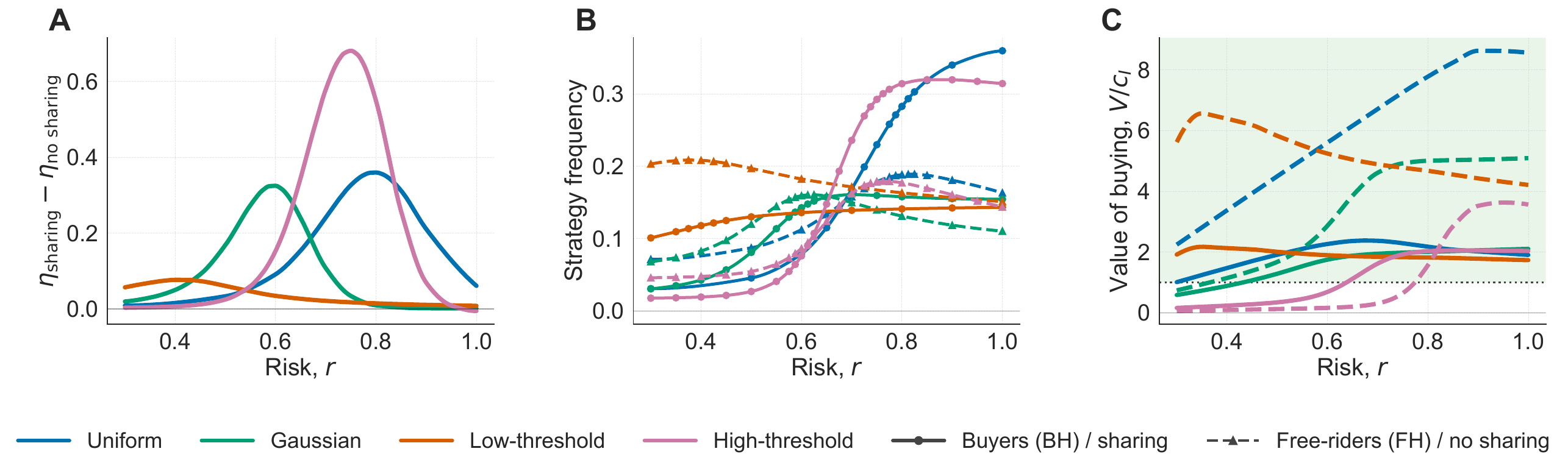}
  \caption{Effect of allowing individuals to share information within their group as a function of risk. \textbf{(A)} Change in group
  success ($\eta$) when information bought by a group member is shared, relative to
  the same five strategies without sharing. \textbf{(B)} Stationary frequency
  of buyers (BH, solid) and information free-riders (FH, dashed) with
  sharing. \textbf{(C)} Value of buying the threshold relative to its price,
  $V/c_I$, for a random individual in the stationary population, with sharing
  (solid) and without (dashed). Buying is privately beneficial in the green region ($V/c_I>1$). ($Z=100$, $\beta=3$, $\mu=0.01$, $N=8$, $c=0.1$,
  $c_I=0.01$, $b=1$, $M_f=4$)}
  \label{fig:shared-info}
\end{figure*}

\paragraph{Sharing turns information into a collective good.}
In Figure~\ref{fig:shared-info} we show how allowing individuals that acquire information to share it within their group changes the previous dynamics. First, the transition to cooperative outcomes (the cooperative tipping point) happens at lower risk, and the gain in group success is maximized at that transition (Figure~\ref{fig:shared-info}A). The gain is also largest for the high-threshold prior, and decreases again at high risk, where groups also succeed without shared information. This happens because a single informed member now is able to coordinate the conditional contributions of the whole group, so information acquired privately raises the success of others.

Figure~\ref{fig:shared-info}B shows that buyers and information free-riders coexist. Free-riders lack information whenever no group member bought it and then defect, so they contribute less often than buyers and, in almost all cases, make fewer pivotal contributions. Also, buyers under uniform and high-threshold priors increase with risk. This is because, under these two priors, high thresholds are common, so a group succeeds only if its conditional cooperators are informed, and as risk increases, the value of buying grows (Figure~\ref{fig:shared-info}C) and buyers come to earn more than the free-riders who rely on them (SI, Figure~\ref{fig:si-shared-info-payoff-gap}).

Sharing also changes the private value of buying (Figure~\ref{fig:shared-info}C). Because, in this setting, information is worthless to a buyer whose co-player already bought it, sharing lowers the value of buying for the uniform, gaussian, and low-threshold priors, but it is still privately beneficial for most of the risk range. Additionally, whether buying impacts payoff negatively depends on risk (SI, Figure~\ref{fig:si-shared-info-payoff-gap}). Below the tipping point, buyers earn less than the free-riders who use the information they share, so they carry a net private cost for the group's success, and buying can only pay when no one else buys. These dynamics are similar to those in a volunteer's dilemma \citep{Diekmann1985}, and each individual prefers that someone else acquires the information.

\begin{figure*}[t!]
  \centering
  \includegraphics[width=0.99\textwidth]{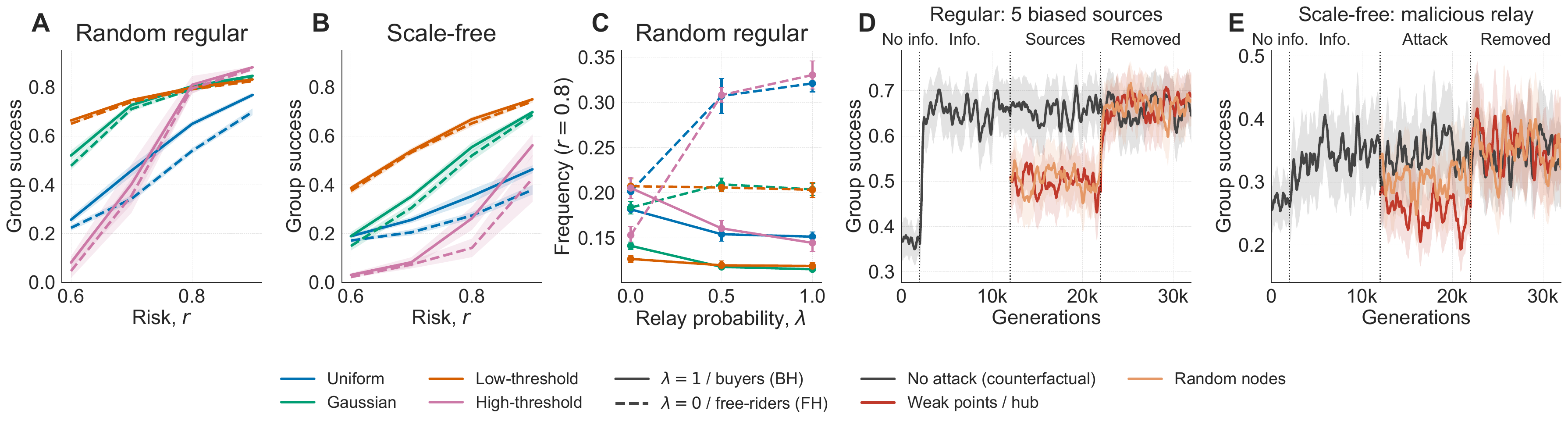}
  \caption{Information diffusion on networks ($Z=100$). \textbf{(A, B)} Group success without relaying ($\lambda=0$, dashed) and with relaying ($\lambda=1$, solid) on random regular (A) and scale-free (B) networks. \textbf{(C)} Frequency of buyers (BH, solid) and information free-riders (FH, dashed) as a function of the relay probability ($r=0.8$, random regular). \textbf{(D, E)} Ensemble-mean group success for the uniform prior ($r=0.8$, $\lambda=1$) over four phases separated by dotted lines: no information strategies, information strategies introduced, attack, and its removal. (D) Five biased sources reporting $M=0$ at weak points (red) or random nodes (orange) on the random regular network. (E) a malicious relay at the hub (red) or at a random node (orange) on the scale-free network. The black ensemble is the counterfactual without attack. Other attacks and priors, and full account of parameters, are summarized in SI. ($\beta=3$, $\mu=0.01$, $c_I=0.01$, $c=0.1$, $b=1$, $M_f=4$)}
  \label{fig:network}
\end{figure*}

\begin{figure}[t]
  \centering
  \includegraphics[width=\linewidth]{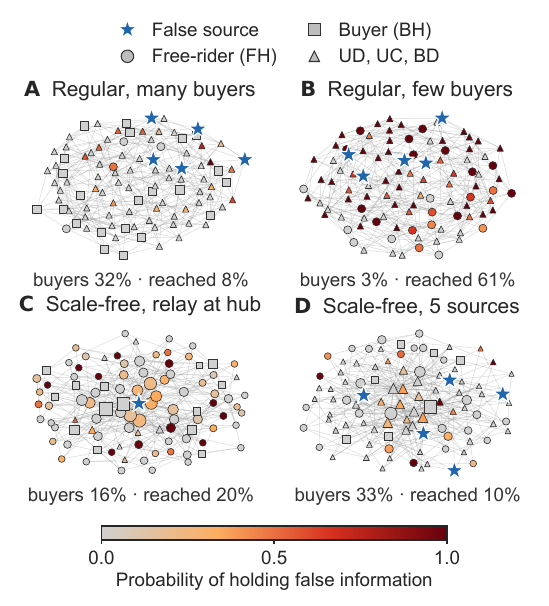}
  \caption{Network snapshots at the end of an attack (uniform prior). Node
  colour is the probability of holding false information (grey - never reached, dark red - always reached), shapes mark information free-riders (circles), buyers (squares),
  and the other strategies (triangles), and blue stars the false sources. On
  scale-free networks, node size grows with degree. (A, B) The same attack, five
  biased sources on the random regular network, with many and with few buyers;
  (C) a malicious relay at a hub and (D) five biased sources at random nodes
  on the scale-free network. The share of buyers is not set by experiment, but emerges from the dynamics.
  Probabilities are averaged over thresholds and 64 cascades. ($Z=100$, $r=0.8$,
  $\lambda=1$, $\beta=3$, $\mu=0.01$, $c=0.1$, $c_I=0.01$, $b=1$, $M_f=4$;
  random 7-regular and Barab\'asi--Albert networks with $m=4$)}
  \label{fig:snapshots}
\end{figure}

\paragraph{Diffusion reduces the number of individuals who pay for information, increases success, and makes the system vulnerable to targeted attacks.}
In Figure~\ref{fig:network} we show how allowing information to diffuse on a network, instead of only within a group has a significant impact on the dynamics. In the absence of relaying ($\lambda=0$), only the neighbours of buyers are informed (similar to the previous well-mixed scenario). Yet, group success is only slightly lower than when information is always relayed ($\lambda=1$), except under the uniform prior on random regular networks and the high-threshold prior on scale-free networks (Figure~\ref{fig:network}A--B). However, Figure~\ref{fig:network}C shows that at $r=0.8$, under the uniform and high-threshold priors, the share of buyers decreases significantly from $\lambda=0$ to $\lambda=1$, while the share of information free-riders increases. This shows that diffusion spreads information widely but shifts its acquisition to fewer individuals. Furthermore, noisy relays do not affect the results significantly and moderate noise even raises success under the uniform prior (SI, Figure~\ref{fig:si-network-noise}).

Diffusion poses an additional problem related to the trust on the source of information. In Figures~\ref{fig:network}D--E we show that malicious relays and external biased sources reduce group success while present, and that the damage is not larger when they are placed at the weak points of the network, where information free-riders concentrate, but it is larger when they are placed at the hubs of scale-free networks. However, the damage is not permanent, and success recovers after their removal (see SI, Section~\ref{sec:si-network-results} for details on other attacks, including large attacks that bring collective action close to collapse).

\paragraph{Information buyers act as a firewall against the spread of false information.}
False information mainly reaches the direct neighbours of its sources and changes the actions only of information free-riders, the only individuals whose contribution depends on what they hear (see Figure~\ref{fig:snapshots} and SI, Section~\ref{sec:si-network-results}). How far it spreads depends on how many individuals buy information, because every buyer starts a cascade of true information that competes with the false one. In this way, the fewer the buyers, the farther disinformation spreads, and the number of buyers acts as a firewall against disinformation. On degree-heterogeneous networks, success is lower overall, but relaying still raises it. However, the placement of sources of false information in this case matters, and false information injected at hubs is considerably more damaging than at random nodes, and a few biased hubs can suppress success almost entirely under the high-threshold prior (Figure~\ref{fig:network}E and SI, Section~\ref{sec:si-network-results}).

\section{Discussion and Conclusion}

We studied whether allowing agents to purchase exact information about an uncertain collective threshold improves cooperation, group success, and welfare in an evolutionary collective-risk dilemma. We find that (i) collective risk and the threshold prior shape the coordination problem even before information acquisition is available; (ii) costly information can improve collective outcomes when sufficiently cheap ($c_I<\varphi c\leq c$); (iii) when information is shared within groups, its acquisition becomes a collective good that moves the cooperation tipping point to lower risk; and (iv) on networks, diffusion spreads information widely but shifts its acquisition to fewer individuals and is robust to noise, while false information harms collective success mostly while present.

\paragraph{Information pays only when cheap, and sharing it is a collective good.}
Threshold uncertainty is known to reduce and destabilize collective action \citep{BarrettDannenberg2014,Dannenberg2015}. Yet, we show here that it need not be resolved only through an external intervention, and can instead be resolved endogenously. We find that individuals acquire information and raise group success above the level reached without information, where collective risk alone already drives a transition to cooperation at a critical risk that depends on the prior \citep{SantosPacheco2011}. Its price, however, is decisive. In line with information-value theory \citep{Howard1966}, information is valuable to an individual only if it can change its decision, that is, only if contributing is worth its cost at some thresholds and not at others. This bounds the value of information by the cost of cooperation itself (Equation~\eqref{eq:voi-bound}), so information is worth buying only when it is considerably cheaper than contributing. Moreover, the simple conditional rules we consider capture none of this private value, because they respond to the threshold rather than to whether their contribution is pivotal. Yet, the contributions they make on the basis of information are jointly decisive for many successful groups, which indicates that information acquired and used in this way is still collectively valuable, even when it impacts negatively the payoff of buyers. Therefore, as with public information in coordination problems \citep{MorrisShin2002,MyattWallace2012}, its social value and the private incentive to acquire it do not need to coincide, which is what makes sharing it consequential.

In this latter case, when information bought by one group member reaches the others, a single informed individual coordinates the contributions of its whole group which shifts the cooperation tipping point towards lower levels of risk. On networks, diffusion makes information available beyond the group in which it was bought, but it also lets more individuals rely on others' purchases, which, as we observe, reduces the number of buyers.

\paragraph{False information harms mostly while present, with the most impact from hubs.}
Consistent with experiments in which locally noisy agents improved global coordination in networks \citep{ShiradoChristakis2017}, noisy relaying did not reduce success in our model. Yet,
false information did, and large attacks brought collective action close to
collapse. The damage lasted mostly while false information was present, and when the sources of false information are removed, success is restored within a few thousand generations even after an attack that had nearly suppressed it. We also find that false information injected at hubs was up to three times more damaging than at random nodes, and that its reach depended on the individuals who buy information, whose cascades of true information act as a firewall against false sources. Acquiring information is therefore a collective good since it protects the population against disinformation. At the same time, the information free-riding that diffusion encourages weakens that protection. 

Here, buyers can also be interpreted as individuals who invest effort in verifying information and correcting others, for instance by flagging misleading posts, on whom platforms rely to reduce misinformation \citep{TreenWilliamsONeill2020}. Such efforts may be particularly relevant in networks where misinformation is concentrated. For example, online discussions of climate change are organized into communities, many of which function as echo chambers \citep{WilliamsEtAl2015}. Moreover, fake news is shared by a very small fraction of accounts \citep{GrinbergEtAl2019}, while corporately funded organizations in the climate change countermovement occupy the most central positions in its network \citep{Farrell2016}. These findings have motivated proposals to monitor the small number of accounts responsible for most fake news \citep{GrinbergEtAl2019} and to detect malicious accounts early \citep{TreenWilliamsONeill2020}. Consistent with the potential benefits of such interventions, we find that success recovers once false sources are removed, suggesting that efforts to identify them can yield lasting benefits.

Overall, our results show that the value of costly information in collective-risk dilemmas is conditional. Returning to our two central questions, acquiring information about an uncertain threshold helps groups avoid collective disaster only when its cost is low relative to cooperation. When information is shared within groups or across networks, a few buyers can benefit many, albeit at the expense of increased free-riding. Importantly, buyers also enhance resilience to misinformation and disinformation by limiting the spread of false information. We find that noise in relayed information does not undermine collective success and that populations fully recover once false sources are removed. Thus, the contribution of information to collective action depends on its cost, the continued willingness of individuals to acquire it, and the reliability of its transmission.

\section{Contributions}
E.F.D.: Conceptualization, methodology, software, formal analysis, investigation, and writing – original draft. T.L.: Conceptualization, supervision and writing – review and editing. G.I. and W.T.: Conceptualization. All authors have read and approved the final manuscript.

\section{Acknowledgements}
 The authors gratefully acknowledge the research support of the Japan Society for the Promotion of Science (JSPS) and the Fonds de la Recherche Scientifique (FNRS) through the Japan-Belgium Bilateral Research Cooperative Program between JSPS and FNRS (Grant Numbers JPJSBP120242201 and 40022755). E.F.D. is supported by an F.W.O. Senior postdoctoral grant (12A7825N). T.L. further acknowledges the support of the Service Public de Wallonie Recherche (grant 2010235-ARIAC) by DigitalWallonia4.ai and the Flemish Government through the AI Research Program. 

\bibliographystyle{plainnat}
\bibliography{references}

@techreport{IPCC2018,
  author = "{IPCC}",
  title  = "Global Warming of 1.5{\textdegree}C. An IPCC Special Report on the Impacts of Global Warming of 1.5{\textdegree}C above Pre-Industrial Levels and Related Global Greenhouse Gas Emission Pathways, in the Context of Strengthening the Global Response to the Threat of Climate Change, Sustainable Development, and Efforts to Eradicate Poverty",
  editor = "Masson-Delmotte, V. and Zhai, P. and P{\"o}rtner, H.-O. and Roberts, D. and Skea, J. and Shukla, P.R. and Pirani, A. and Moufouma-Okia, W. and P{\'e}an, C. and Pidcock, R. and Connors, S. and Matthews, J.B.R. and Chen, Y. and Zhou, X. and Gomis, M.I. and Lonnoy, E. and Maycock, T. and Tignor, M. and Waterfield, T.",
  year   = 2018,
  institution = "Intergovernmental Panel on Climate Change",
  doi    = {10.1017/9781009157940},
}

@article{Lazer2018,
  author  = "Lazer, David M. J. and Baum, Matthew A. and Benkler, Yochai and Berinsky, Adam J. and Greenhill, Kelly M. and Menczer, Filippo and Metzger, Miriam J. and Nyhan, Brendan and Pennycook, Gordon and Rothschild, David and Schudson, Michael and Sloman, Steven A. and Sunstein, Cass R. and Thorson, Emily A. and Watts, Duncan J. and Zittrain, Jonathan L.",
  title   = "The Science of Fake News",
  journal = "Science",
  volume  = 359,
  number  = 6380,
  pages   = "1094--1096",
  year    = 2018,
  doi     = "10.1126/science.aao2998",
}

@article{Nagler2020,
  author  = "Nagler, Rebekah H. and Vogel, Rachel I. and Gollust, Sarah E. and Rothman, Alexander J. and Fowler, Erika Franklin and Yzer, Marco C.",
  title   = "Public Perceptions of Conflicting Information Surrounding {COVID}-19: Results from a Nationally Representative Survey of {U.S.} Adults",
  journal = "PLOS ONE",
  volume  = 15,
  number  = 10,
  pages   = "e0240776",
  year    = 2020,
  doi     = "10.1371/journal.pone.0240776",
}

@article{Meng2023,
  author  = "Meng, Yao and Broom, Mark and Li, Aming",
  title   = "Impact of Misinformation in the Evolution of Collective Cooperation on Networks",
  journal = "Journal of the Royal Society Interface",
  volume  = 20,
  number  = 206,
  pages   = "20230295",
  year    = 2023,
  doi     = "10.1098/rsif.2023.0295",
}

@inproceedings{ChenShu2024,
  author    = "Chen, Canyu and Shu, Kai",
  title     = "Can {LLM}-Generated Misinformation Be Detected?",
  booktitle = "Proceedings of the Twelfth International Conference on Learning Representations (ICLR)",
  year      = 2024,
  url       = "https://openreview.net/forum?id=ccxD4mtkTU",
  eprint    = "2309.13788",
  archivePrefix = "arXiv",
}

@article{Milinski2008,
  author  = "Milinski, Manfred and Sommerfeld, Ralf D. and Krambeck, Hans-J{\"u}rgen and Reed, Floyd A. and Marotzke, Jochem",
  title   = "The Collective-Risk Social Dilemma and the Prevention of Simulated Dangerous Climate Change",
  journal = "Proceedings of the National Academy of Sciences",
  volume  = 105,
  number  = 7,
  pages   = "2291--2294",
  year    = 2008,
  doi     = "10.1073/pnas.0709546105",
}

@article{SantosPacheco2011,
  author  = "Santos, Francisco C. and Pacheco, Jorge M.",
  title   = "Risk of Collective Failure Provides an Escape from the Tragedy of the Commons",
  journal = "Proceedings of the National Academy of Sciences",
  volume  = 108,
  number  = 26,
  pages   = "10421--10425",
  year    = 2011,
  doi     = "10.1073/pnas.1015648108",
}

@article{TavoniEtAl2011,
  author  = "Tavoni, Alessandro and Dannenberg, Astrid and Kallis, Giorgos and L{\"o}schel, Andreas",
  title   = "Inequality, Communication, and the Avoidance of Disastrous Climate Change in a Public Goods Game",
  journal = "Proceedings of the National Academy of Sciences",
  volume  = 108,
  number  = 29,
  pages   = "11825--11829",
  year    = 2011,
  doi     = "10.1073/pnas.1102493108",
}

@article{BarrettDannenberg2012,
  author  = "Barrett, Scott and Dannenberg, Astrid",
  title   = "Climate Negotiations under Scientific Uncertainty",
  journal = "Proceedings of the National Academy of Sciences",
  volume  = 109,
  number  = 43,
  pages   = "17372--17376",
  year    = 2012,
  doi     = "10.1073/pnas.1208417109",
}

@article{BarrettDannenberg2014,
  author  = "Barrett, Scott and Dannenberg, Astrid",
  title   = "Sensitivity of Collective Action to Uncertainty about Climate Tipping Points",
  journal = "Nature Climate Change",
  volume  = 4,
  number  = 1,
  pages   = "36--39",
  year    = 2014,
  doi     = "10.1038/nclimate2059",
}

@article{Dannenberg2015,
  author  = "Dannenberg, Astrid and L{\"o}schel, Andreas and Paolacci, Gabriele and Reif, Christiane and Tavoni, Alessandro",
  title   = "On the Provision of Public Goods with Probabilistic and Ambiguous Thresholds",
  journal = "Environmental and Resource Economics",
  volume  = 61,
  number  = 3,
  pages   = "365--383",
  year    = 2015,
  doi     = "10.1007/s10640-014-9796-6",
}

@article{PachecoVasconcelosSantos2014,
  author  = "Pacheco, Jorge M. and Vasconcelos, V{\'i}tor V. and Santos, Francisco C.",
  title   = "Climate Change Governance, Cooperation and Self-Organization",
  journal = "Physics of Life Reviews",
  volume  = 11,
  number  = 4,
  pages   = "573--586",
  year    = 2014,
  doi     = "10.1016/j.plrev.2014.02.003",
}

@article{FernandezDomingos2020,
  author  = "Fern{\'a}ndez Domingos, Elias and Gruji{\'c}, Jelena and Burguillo, Juan C. and Kirchsteiger, Georg and Santos, Francisco C. and Lenaerts, Tom",
  title   = "Timing Uncertainty in Collective Risk Dilemmas Encourages Group Reciprocation and Polarization",
  journal = "iScience",
  volume  = 23,
  number  = 12,
  pages   = "101752",
  year    = 2020,
  doi     = "10.1016/j.isci.2020.101752",
}

@article{FernandezDomingos2023,
  author  = "Fern{\'a}ndez Domingos, Elias and Santos, Francisco C. and Lenaerts, Tom",
  title   = "{EGT}tools: Evolutionary Game Dynamics in Python",
  journal = "iScience",
  volume  = 26,
  number  = 4,
  pages   = "106419",
  year    = 2023,
  doi     = "10.1016/j.isci.2023.106419",
}

@article{RolnickEtAl2022,
  author  = "Rolnick, David and Donti, Priya L. and Kaack, Lynn H. and Kochanski, Kelly and Lacoste, Alexandre and Sankaran, Kris and Ross, Andrew Slavin and Milojevic-Dupont, Nikola and Jaques, Natasha and Waldman-Brown, Anna and Luccioni, Alexandra Sasha and Maharaj, Tegan and Sherwin, Evan D. and Mukkavilli, S. Karthik and Kording, Konrad P. and Gomes, Carla P. and Ng, Andrew Y. and Hassabis, Demis and Platt, John C. and Creutzig, Felix and Chayes, Jennifer and Bengio, Yoshua",
  title   = "Tackling Climate Change with Machine Learning",
  journal = "ACM Computing Surveys",
  volume  = 55,
  number  = 2,
  pages   = "42",
  year    = 2023,
  doi     = "10.1145/3485128",
}

@misc{ShiWangFang2020,
  author        = "Shi, Zheyuan Ryan and Wang, Claire and Fang, Fei",
  title         = "Artificial Intelligence for Social Good: A Survey",
  year          = 2020,
  eprint        = "2001.01818",
  archivePrefix = "arXiv",
  doi           = {10.48550/arXiv.2001.01818},
  url           = {https://arxiv.org/abs/2001.01818},
}

@inproceedings{ZhangEtAl2025,
  author    = "Zhang, Tianyu and Williams, Andrew Robert and Wozny, Phillip and Cohrs, Kai-Hendrik and Ponse, Koen and Jiralerspong, Marco and Phade, Soham Rajesh and Srinivasa, Sunil and Li, Lu and Zhang, Yang and Gupta, Prateek and Acar, Erman and Rish, Irina and Bengio, Yoshua and Zheng, Stephan",
  title     = "{AI} for Global Climate Cooperation: Modeling Global Climate Negotiations, Agreements, and Long-Term Cooperation in {RICE-N}",
  booktitle = "Proceedings of the 42nd International Conference on Machine Learning (ICML)",
  series    = "Proceedings of Machine Learning Research",
  volume    = 267,
  publisher = "PMLR",
  url       = "https://proceedings.mlr.press/v267/zhang25ce.html",
  pages     = "76332--76360",
  year      = 2025,
}

@article{Merhej2022,
  author  = "Merhej, Ramona and Santos, Fernando P. and Melo, Francisco S. and Santos, Francisco C.",
  title   = "Cooperation and Learning Dynamics under Wealth Inequality and Diversity in Individual Risk",
  journal = "Journal of Artificial Intelligence Research",
  volume  = 74,
  pages   = "733--764",
  year    = 2022,
  doi     = "10.1613/jair.1.13519",
}

@article{MyattWallace2012,
  author  = "Myatt, David P. and Wallace, Chris",
  title   = "Endogenous Information Acquisition in Coordination Games",
  journal = "Review of Economic Studies",
  volume  = 79,
  number  = 1,
  pages   = "340--374",
  year    = 2012,
  doi     = "10.1093/restud/rdr018",
}

@article{MorrisShin2002,
  author  = "Morris, Stephen and Shin, Hyun Song",
  title   = "Social Value of Public Information",
  journal = "American Economic Review",
  volume  = 92,
  number  = 5,
  pages   = "1521--1534",
  year    = 2002,
  doi     = "10.1257/000282802762024610",
}

@article{VanDriesscheEtAl2024,
  author  = "Van Driessche, Sarah and Boun My, Kene and Brunette, Marielle",
  title   = "Risk and Ambiguity in a Public Good Game",
  journal = "Frontiers in Behavioral Economics",
  volume  = 3,
  pages   = "1456436",
  year    = 2024,
  doi     = "10.3389/frbhe.2024.1456436",
}

@misc{XuStangenberg2025,
  author = "Xu, Yan and Stangenberg, Lennart",
  title  = "You Know, So I Know! Endogenous Information Acquisition and Observability in a Threshold Public Goods Game",
  year   = 2025,
  doi    = "10.2139/ssrn.5651650",
  howpublished = {SSRN working paper 5651650},
}

@book{HofbauerSigmund1998,
  author    = "Hofbauer, Josef and Sigmund, Karl",
  title     = "Evolutionary Games and Population Dynamics",
  publisher = "Cambridge University Press",
  address   = "Cambridge",
  year      = 1998,
  doi       = {10.1017/CBO9781139173179},
}

@article{TraulsenNowakPacheco2006,
  author  = "Traulsen, Arne and Nowak, Martin A. and Pacheco, Jorge M.",
  title   = "Stochastic Dynamics of Invasion and Fixation",
  journal = "Physical Review E",
  volume  = 74,
  number  = 1,
  pages   = "011909",
  year    = 2006,
  doi     = "10.1103/PhysRevE.74.011909",
}

@article{CoutoPachecoSantos2020,
  author  = "Couto, Marta C. and Pacheco, Jorge M. and Santos, Francisco C.",
  title   = "Governance of Risky Public Goods under Graduated Punishment",
  journal = "Journal of Theoretical Biology",
  volume  = 505,
  pages   = "110423",
  year    = 2020,
  doi     = "10.1016/j.jtbi.2020.110423",
}

@article{VasconcelosSantosPacheco2013,
  author  = "Vasconcelos, V{\'i}tor V. and Santos, Francisco C. and Pacheco, Jorge M.",
  title   = "A Bottom-Up Institutional Approach to Cooperative Governance of Risky Commons",
  journal = "Nature Climate Change",
  volume  = 3,
  number  = 9,
  pages   = "797--801",
  year    = 2013,
  doi     = "10.1038/nclimate1927",
}

@article{Howard1966,
  author  = "Howard, Ronald A.",
  title   = "Information Value Theory",
  journal = "IEEE Transactions on Systems Science and Cybernetics",
  volume  = 2,
  number  = 1,
  pages   = "22--26",
  year    = 1966,
  doi     = "10.1109/TSSC.1966.300074",
}

@article{garrido2025new,
  title={A new electoral bottom-up model of institutional governance},
  author={Garrido, Carlos M and Santos, Francisco C and Fern{\'a}ndez Domingos, Elias and Nunes, Ana M and Pacheco, Jorge M},
  journal={Scientific Reports},
  volume={15},
  number={1},
  pages={3865},
  year={2025},
  doi={10.1038/s41598-025-87322-y},
  publisher={Nature Publishing Group UK London}
}

@article{ShiradoChristakis2017,
  author       = {Shirado, Hirokazu and Christakis, Nicholas A.},
  title        = {Locally noisy autonomous agents improve global human coordination in network experiments},
  journal      = {Nature},
  volume       = {545},
  number       = {7654},
  pages        = {370--374},
  year         = {2017},
  doi          = {10.1038/nature22332},
}

@article{OhtsukiEtAl2006,
  author       = {Ohtsuki, Hisashi and Hauert, Christoph and Lieberman, Erez and Nowak, Martin A.},
  title        = {A simple rule for the evolution of cooperation on graphs and social networks},
  journal      = {Nature},
  volume       = {441},
  number       = {7092},
  pages        = {502--505},
  year         = {2006},
  doi          = {10.1038/nature04605},
}

@article{SantosSantosPacheco2008,
  author       = {Santos, Francisco C. and Santos, Marta D. and Pacheco, Jorge M.},
  title        = {Social diversity promotes the emergence of cooperation in public goods games},
  journal      = {Nature},
  volume       = {454},
  number       = {7201},
  pages        = {213--216},
  year         = {2008},
  doi          = {10.1038/nature06940},
}

@article{Centola2010,
  author       = {Centola, Damon},
  title        = {The Spread of Behavior in an Online Social Network Experiment},
  journal      = {Science},
  volume       = {329},
  number       = {5996},
  pages        = {1194--1197},
  year         = {2010},
  doi          = {10.1126/science.1185231},
}

@article{VosoughiRoyAral2018,
  author       = {Vosoughi, Soroush and Roy, Deb and Aral, Sinan},
  title        = {The spread of true and false news online},
  journal      = {Science},
  volume       = {359},
  number       = {6380},
  pages        = {1146--1151},
  year         = {2018},
  doi          = {10.1126/science.aap9559},
}

@article{Diekmann1985,
  author       = {Diekmann, Andreas},
  title        = {Volunteer's Dilemma},
  journal      = {Journal of Conflict Resolution},
  volume       = {29},
  number       = {4},
  pages        = {605--610},
  year         = {1985},
  doi          = {10.1177/0022002785029004003},
}

@article{Farrell2016,
  author       = {Farrell, Justin},
  title        = {Corporate funding and ideological polarization about climate change},
  journal      = {Proceedings of the National Academy of Sciences},
  volume       = {113},
  number       = {1},
  pages        = {92--97},
  year         = {2016},
  doi          = {10.1073/pnas.1509433112},
}

@article{GrinbergEtAl2019,
  author       = {Grinberg, Nir and Joseph, Kenneth and Friedland, Lisa and Swire-Thompson, Briony and Lazer, David},
  title        = {Fake news on {Twitter} during the 2016 {U.S.} presidential election},
  journal      = {Science},
  volume       = {363},
  number       = {6425},
  pages        = {374--378},
  year         = {2019},
  doi          = {10.1126/science.aau2706},
}

@article{TreenWilliamsONeill2020,
  author       = {Treen, Kathie M. d'I. and Williams, Hywel T. P. and O'Neill, Saffron J.},
  title        = {Online misinformation about climate change},
  journal      = {WIREs Climate Change},
  volume       = {11},
  number       = {5},
  pages        = {e665},
  year         = {2020},
  doi          = {10.1002/wcc.665},
}

@article{WilliamsEtAl2015,
  author       = {Williams, Hywel T. P. and McMurray, James R. and Kurz, Tim and Lambert, F. Hugo},
  title        = {Network analysis reveals open forums and echo chambers in social media discussions of climate change},
  journal      = {Global Environmental Change},
  volume       = {32},
  pages        = {126--138},
  year         = {2015},
  doi          = {10.1016/j.gloenvcha.2015.03.006},
}

@article{pacheco2026revenue,
  title={Revenue-feedback and the institutional governance of global risky commons},
  author={Pacheco, Jorge M and Garrido, Carlos M and Fern{\'a}ndez Domingos, Elias},
  journal={Journal of The Royal Society Interface},
  volume={23},
  number={241},
  pages={20260508},
  year={2026},
  doi={10.1098/rsif.2026.0508},
  publisher={The Royal Society}
}

@article{SzaboFath2007,
  author       = {Szab{\'o}, Gy{\"o}rgy and F{\'a}th, G{\'a}bor},
  title        = {Evolutionary games on graphs},
  journal      = {Physics Reports},
  volume       = {446},
  number       = {4--6},
  pages        = {97--216},
  year         = {2007},
  doi          = {10.1016/j.physrep.2007.04.004},
}

\clearpage
\appendix
\raggedbottom
\renewcommand{\thesection}{S\arabic{section}}
\renewcommand{\thefigure}{S\arabic{figure}}
\renewcommand{\thetable}{S\arabic{table}}
\renewcommand{\theequation}{S\arabic{equation}}
\renewcommand{\thealgorithm}{S\arabic{algorithm}}
\setcounter{figure}{0}
\setcounter{table}{0}
\setcounter{equation}{0}
\setcounter{algorithm}{0}

\twocolumn[
  \begin{@twocolumnfalse}
    \begin{center}
      {\LARGE\bfseries Supporting Information for:}\par\vspace{1em}
      {\LARGE\bfseries The Conditional Value of Costly Information in Collective-Risk Dilemmas}\\
    {\large Elias Fern\'andez Domingos, Genki Ichinose, Wakaba Tateishi and Tom Lenaerts\par}
    \end{center}
    \vspace{1.5em}
  \end{@twocolumnfalse}
]

\section{Notation}
\label{sec:si-notation}

\begin{figure}[b]
  \centering
  \includegraphics[width=0.75\linewidth]{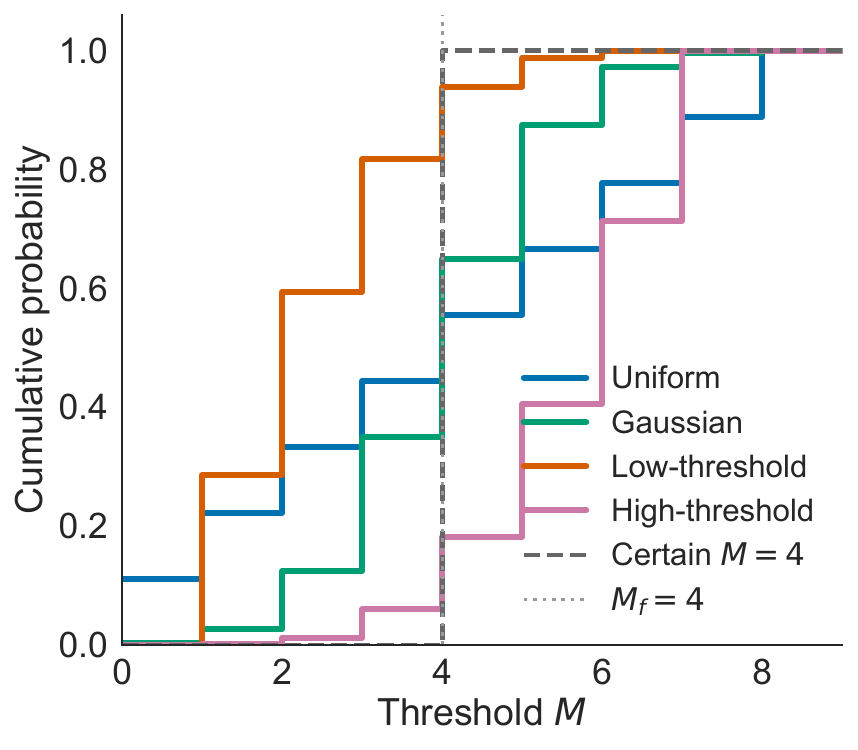}
  \caption{Cumulative distributions of the four threshold priors $p(M)$,
  $M\in\{0,\dots,N\}$, with the certain threshold $M=4$ for comparison.}
  \label{fig:si-discrete-threshold-priors}
\end{figure}

Table~\ref{tab:si-notation} collects the notation. Group size ($N=8$), population size ($Z=100$), and cooperation cost ($c=0.1$) are fixed throughout. The reference threshold $M_f=4$ is identical across priors and it equals the prior mean for the uniform and gaussian priors, but not for the low- and high-threshold priors ($\overline M=2.38$ and $5.62$; Figure~\ref{fig:si-discrete-threshold-priors}).

\begin{table*}[t]
\centering
\small
\begin{tabular}{@{}l p{0.78\textwidth}@{}}
\toprule
Symbol & Meaning \\
\midrule
$Z$, $N$ & Population size; group size \\
$M$ & Realized threshold (minimum number of cooperators for group success), $M\in\{0,\dots,N\}$ \\
$p(M)$, $\overline{M}$ & Prior probability mass function of the threshold; its mean \\
$r$ & Collective risk: probability of losing the endowment if the group fails \\
$b$ & Endowment; all payoffs are reported in units of $b$ \\
$c$, $c_I$ & Cooperation cost and information cost, as fractions of $b$ \\
$\Sset$ & Strategy set; $\{\mathrm{UD},\mathrm{UC},\mathrm{CH},\mathrm{CL}\}$ unless stated otherwise \\
$M_f$ & Fixed reference threshold of the conditional strategies, independent of $p(M)$; $M_f=4$ \\
$q_i(M)$ & Action of strategy $i$ at threshold $M$ ($1$ = contribute, $0$ = defect) \\
$I_i$ & Information-purchase indicator of strategy $i$ \\
$\ind\{\cdot\}$, $x^{+}$ & Indicator function; $\max\{x,0\}$ \\
$\Prb$, $\E_p$ & Probability; expectation over the prior $p(M)$ \\
\midrule
$k$, $j$ & Number of cooperators in the group (focal included); among the focal's $N-1$ co-players \\
$\pi_D(k,M)$, $\pi_C(k,M)$ & Payoff of a defector and of a cooperator in a group with $k$ cooperators \\
$\pi_i(j,M)$ & Payoff of a focal $i$ facing $j$ cooperating co-players \\
$\Pi_i$, $\overline\Pi$ & Expected payoff (fitness) of strategy $i$; population-average payoff \\
\midrule
$x_i$, $\mathbf{x}$ & Frequency of strategy $i$ in an infinite population; frequency vector \\
$\overline q(M)$ & Probability that a random co-player contributes at threshold $M$, $\sum_{\ell}x_\ell q_\ell(M)$ \\
$K_i$, $\mathbf{K}$ & Number of individuals using strategy $i$ in a finite population; population state, $\sum_iK_i=Z$ \\
$\kappa_i(M)$, $\kappa(M)$ & Contributors at threshold $M$ among the $Z-1$ others of a focal $i$; in the whole population \\
$\beta$, $\mu$ & Intensity of selection; mutation (exploration) probability \\
$\rho(\mathbf{K})$ & Stationary distribution of the finite-population Markov chain \\
\midrule
$\eta$ & Group success: probability that a random group reaches its threshold \\
$f_{\rm coop}$, $f_{\rm info}$ & Cooperation frequency; information-purchase frequency \\
$f$, $f^{*}$ & Cooperation frequency in the two-strategy model ($=x_{\rm UC}$); its payoff-maximizing value \\
\midrule
$\mathrm{Piv}(M)$ & Probability that exactly $M-1$ co-players contribute, so the focal contribution is pivotal \\
$\Delta\Pi(M)$ & Expected payoff gain from contributing rather than defecting at threshold $M$ \\
$V$, $V_i$ & Value of information; value realized by strategy $i$ \\
$\theta(M)$, $\varphi$ & Pivotality-to-cost ratio $r\,\mathrm{Piv}(M)/c$; value of information in units of $cb$ \\
\midrule
BH, FH, BD & Shared information: buyer, information free-rider, information provider \\
$\lambda$, $\sigma$ & Networks: relay probability; standard deviation of relay noise \\
NI, CI & No-information baseline (UD and UC, each duplicated); costly-information model \\
\bottomrule
\end{tabular}
\caption{Notation.}
\label{tab:si-notation}
\end{table*}

\section{Payoffs, fitness, and value of information}

\paragraph{Two-strategy reduction.} With UD and UC only and $f=x_{\rm UC}$, a contribution matters only when exactly $M-1$ of the $N-1$ co-players cooperate, so that
\begin{equation}
\label{eq:pivotal}
\Pi_{\rm UC}-\Pi_{\rm UD}
=rb\sum_{M=1}^{N}p(M)\binom{N-1}{M-1}f^{M-1}(1-f)^{N-M}-cb,
\end{equation}
which gives the replicator analysis in Figure~\ref{fig:combined-optima}.

\paragraph{Finite-population fitness and population-level variables.}
\label{sec:si-variables}
For a focal individual using strategy $i$, let $\kappa_i(M)=\sum_{\ell\in\Sset}K_\ell q_\ell(M)-q_i(M)$ be the number of contributors at threshold $M$ among the other $Z-1$ individuals. Its $N-1$ co-players are sampled without replacement, so
\begin{equation}
\label{eq:finite-payoff}
\Pi_i(\mathbf K)
=\sum_{M=0}^{N}p(M)\sum_{j=0}^{N-1}
\frac{\binom{\kappa_i(M)}{j}\binom{Z-1-\kappa_i(M)}{N-1-j}}{\binom{Z-1}{N-1}}
\pi_i(j,M).
\end{equation}
With $\kappa(M)=\sum_{\ell}K_\ell q_\ell(M)$ contributors in the whole population, group success is
$\eta(\mathbf K)=\sum_{M}p(M)\sum_{k\geq M}\binom{\kappa(M)}{k}\binom{Z-\kappa(M)}{N-k}/\binom{Z}{N}$, and the main text reports its stationary average $\sum_{\mathbf K}\rho(\mathbf K)\eta(\mathbf K)$. Cooperation and information-purchase frequencies are $f_{\rm coop}=x_{\rm UC}+x_{\rm CH}\Prb(M\geq M_f)+x_{\rm CL}\Prb(M<M_f)$ and $f_{\rm info}=x_{\rm CH}+x_{\rm CL}$, with $x_i=K_i/Z$.

\paragraph{Value of information.}
\label{sec:si-voi-derivation}
We derive Equation~\eqref{eq:marginal-contribution}. Fix a focal individual of type $i$ in state $\mathbf K$, a threshold $M$, and its information status, so that the acquisition cost is the same whichever action it takes. If $j$ of its $N-1$ co-players contribute, Equation~\eqref{eq:primitive-payoff} gives the payoff $b[1-r+r\,\ind\{j+a\geq M\}]-a\,cb-I_ic_Ib$ for action $a\in\{0,1\}$, so contributing instead of defecting changes it by $b\,r[\ind\{j+1\geq M\}-\ind\{j\geq M\}]-cb=b[r\,\ind\{j=M-1\}-c]$. The co-players do not observe the focal's action, so the distribution of $j$ is the same under both actions, and averaging over $j$ gives $\Delta\Pi(M)=b[r\,\mathrm{Piv}(M)-c]$ with
$\mathrm{Piv}(M)=\binom{\kappa_i(M)}{M-1}\binom{Z-1-\kappa_i(M)}{N-M}/\binom{Z-1}{N-1}$ for $1\leq M\leq N$ and $\mathrm{Piv}(0)=0$. An uninformed individual takes one action for all thresholds and gains at most $\max\{\E_p[\Delta\Pi],0\}$ over defection while an informed one chooses per threshold and gains $\E_p[\max\{\Delta\Pi,0\}]$. Equation~\eqref{eq:voi} states their difference. With $\Delta\Pi^{\pm}=\max\{\pm\Delta\Pi,0\}$, $V=\E_p[\Delta\Pi^{+}]-\max\{\E_p[\Delta\Pi^{+}]-\E_p[\Delta\Pi^{-}],0\}
=\min\{\E_p[\Delta\Pi^{+}],\E_p[\Delta\Pi^{-}]\}$, and writing $\Delta\Pi=cb(\theta-1)$ gives $\Delta\Pi^{-}=cb(1-\theta)^{+}\leq cb\,\ind\{\theta<1\}$, hence Equation~\eqref{eq:voi-bound}. In
Figure~\ref{fig:baseline-comparison}C, $V$ and $V_i$ are evaluated in every population state and averaged over individuals and over the stationary distribution.

\section{No-information baseline}
\label{sec:si-ni}

Mutation acts on strategy labels. In a two-strategy game, a mutating defector always becomes a cooperator, whereas with UD and UC duplicated it does so only with probability $2/3$. Adding payoff-identical copies of UD and UC therefore allows us to correctly compare to CI with this four-strategy baseline, and we use the two-strategy game only for the replicator analysis.

\section{Collective pivotality}
\label{sec:si-collective-pivotality}

When several group members act on the same information, their contributions can be jointly decisive without any one of them being individually pivotal. We call a successful group information-pivotal if it
would fail without the contributions of the members acting on information (CH and CL; with sharing, BH and informed FH), and compute this exactly from the stationary distribution of group compositions. At $c_I=0.01$, information is jointly decisive for up to $48$--$96\%$ of successful groups, depending on the prior, two to three times more often than an informed contributor is individually pivotal; with sharing, this rises to up to $88\%$ under the high-threshold prior.

\section{Sharing information within groups}
\label{sec:si-sharing}

Per-strategy averages use the stationary distribution of group compositions,
$w(\mathbf g)=\sum_{\mathbf K}\rho(\mathbf K)\prod_\ell\binom{K_\ell}{g_\ell}/\binom{Z}{N}$, with an individual of strategy $i$ in a group of composition $\mathbf g$ with probability proportional to $w(\mathbf g)\,g_i/N$. The value of buying (Figure~\ref{fig:shared-info}C) is Equation~\eqref{eq:voi} for a random individual given its co-players; with sharing, buying also informs the FH
co-players and has no value when a co-player already buys. Each prior is solved on 12 to 22 risk values, refined around every turning point and interpolated to produce the plotted curves.

\section{Network simulations}
\label{sec:si-network}

\paragraph{Networks.} Random 7-regular graphs and Barab\'asi--Albert graphs ($m=4$, mean degree $7.7$) on $Z=100$ nodes. Every node plays in the group formed by itself and its neighbours and in those centred on its neighbours \citep{SantosSantosPacheco2008}, however, we use here average payoffs instead of accumulated ones. On Barab\'asi--Albert graphs, a group of size $n_g$ has threshold $\lceil M\,n_g/N\rceil$, which keeps the required share of contributors
constant.

\paragraph{Diffusion and attacks.} All groups in the network share the same threshold $M$. Buyers (BH, BD) and any false sources start the cascade in random order and tell all their neighbours. Every informed BH or FH relays what it received to each uninformed neighbour with probability $\lambda$, and
a node keeps the first message it receives. With relay noise, each transmission adds Gaussian noise of standard deviation $\sigma$, rounded to $\{0,\dots,N\}$. Malicious relays and external sources are fixed nodes that never update, are never imitated, and never contribute. Also, malicious relays and biased sources report $M=0$, noisy sources report the true threshold plus noise of standard deviation $2$, and silent sources nothing. They are placed at the node of highest betweenness centrality or degree (hubs), at the nodes whose neighbourhood has the largest share of FH (weak points), or at random, and rejoin the population with a random strategy when removed.

\begin{algorithm}[H]
\caption{Expected payoff of an individual in a given population state.}
\label{alg:payoff}
\begin{algorithmic}[1]
\small
\State Draw $S$ independent diffusion cascades from the current state ($S=1$ if diffusion is deterministic, i.e.\ $\lambda\in\{0,1\}$ without noise or sources; otherwise $S=16$)
\For{each cascade}
  \For{each threshold $M$ with $p(M)>0$}
    \State Each informed BH or FH contributes if the value it received is at least $M_f$; UC always contributes; UD, BD, sources, and uninformed FH do not
    \State Compute the individual's payoff in each of its groups, average over its groups, and subtract $c_I$ if it buys information
  \EndFor
  \State Sum the payoffs over thresholds, weighted by $p(M)$ (exact expectation over $M$)
\EndFor
\State \Return the average over the $S$ cascades (estimated expectation over cascades)
\end{algorithmic}
\end{algorithm}

\begin{algorithm}[H]
\caption{Imitation dynamics on the network (one update).}
\label{alg:update}
\begin{algorithmic}[1]
\small
\State Choose an individual $i$ uniformly at random (sources and malicious relays never update)
\If{with probability $\mu$}
  \State Exploration: $i$ adopts a strategy drawn uniformly from the other available strategies
\Else
  \State Choose a neighbour $j$ of $i$ uniformly at random
  \If{$j$ uses a different strategy}
    \State Estimate the payoffs $\Pi_i$ and $\Pi_j$ with Algorithm~\ref{alg:payoff}, using the same cascades for both
    \State Imitation: $i$ adopts $j$'s strategy with probability $(1+e^{-\beta(\Pi_j-\Pi_i)})^{-1}$
  \EndIf
\EndIf
\end{algorithmic}
\end{algorithm}

\paragraph{Experiments.} One generation is $Z$ updates. Stationary averages (Figure~\ref{fig:network}A--C) are time averages over $3\times10^{4}$ generations after $5\times10^{3}$ generations of burn-in, from 10 chains per point (5 networks, each started from a random state and from full defection). Shock ensembles (Figure~\ref{fig:network}D, E) start from UD and UC only ($10^{4}$ generations of burn-in, followed by the $2\times10^{3}$ generations shown), then allow the information strategies, add an attack, and remove it ($10^{4}$ generations each, $2\times10^{4}$ for recovery after large attacks), with 60 replicates per arm on 10 networks; an arm without attack is the counterfactual. Ensemble curves are smoothed with a centred moving average over a 300-generation window within each phase, and bands and error bars are 95\% intervals across chains or replicates. In the control without information, BH behave as UC and FH and BD as UD. In the malicious-relay experiment on random regular networks, $S=1$ was used during the attack. The full implementation is in the accompanying code.

\section{Additional network results}
\label{sec:si-network-results}

\paragraph{Attacks and placement.} On random regular networks, a malicious relay lowers success by $0.07$--$0.08$ under the uniform prior and by at most $0.06$ under the other priors; one biased source lowers it by $0.04$--$0.06$ and five by $0.14$--$0.16$. Weak points, where FH make up $0.60$--$0.72$ of
the neighbourhood, are no more vulnerable than random nodes (differences of at most $0.035$). On Barab\'asi--Albert networks, malicious relays and biased sources at hubs lower success $1.5$ to $3$ times as much as at random nodes, and five biased sources at hubs take high-threshold success from $0.27$ to $0.02$. Success returns to the counterfactual within 20 generations after a relay is removed and within about $10^{3}$ generations after five sources.

\paragraph{Large attacks.} Under the high-threshold prior on random regular networks, $20$ biased sources lower success from $0.81$ to $0.12$, against $0.41$ for $20$ silent sources; after their removal, success stays $0.35$--$0.50$ below the counterfactual for about $10^{3}$ generations (three
to four times the deficit after silent sources) and recovers within $5\times10^{3}$ generations.

\begin{figure*}[t]
  \centering
  \includegraphics[width=0.98\textwidth]{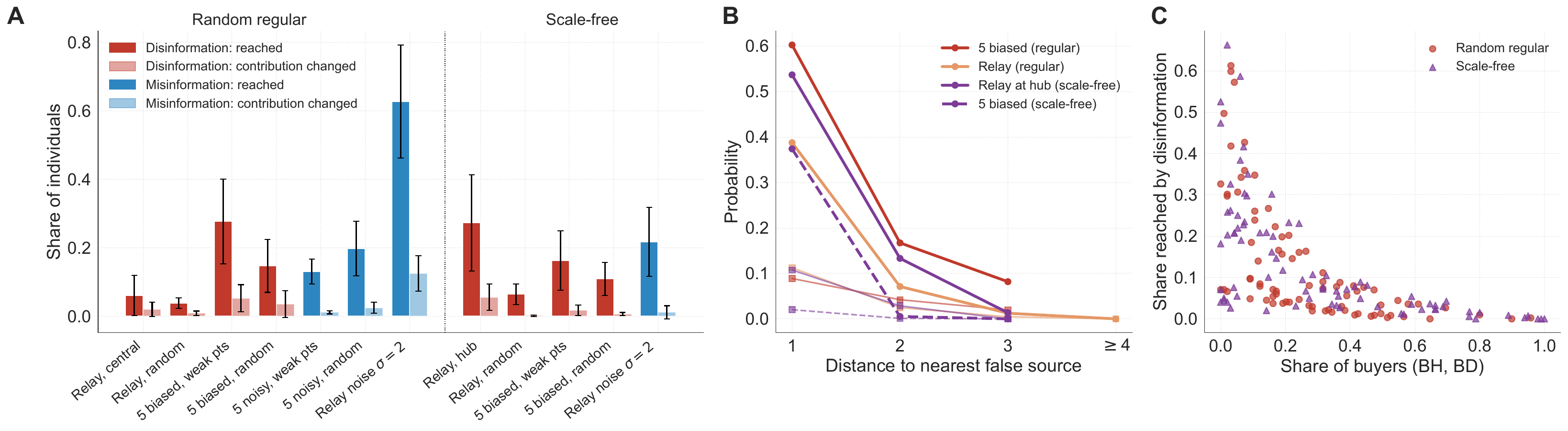}
  \caption{Spread of false information at the end of an attack (uniform
  prior, $r=0.8$, $\lambda=1$). \textbf{(A)} Share of individuals reached by
  disinformation (red) or misinformation (blue), dark, and share whose
  contribution changes, light; 95\% intervals across 10 replicates per
  attack, each taken at the end of the $10^{4}$-generation attack phase,
  and across 6 chains for relay noise ($2\times10^{4}$ generations after
  $5\times10^{3}$ generations of burn-in).
  \textbf{(B)} Probability of being reached (circles) and of changing
  contribution (squares) with the distance to the nearest false source.
  \textbf{(C)} Share reached by disinformation against the share of buyers in
  each end-of-attack state, for the uniform and high-threshold priors.
  Probabilities are averaged over thresholds and 32 cascades per threshold.
  ($Z=100$, $\beta=3$, $\mu=0.01$, $c=0.1$, $c_I=0.01$, $b=1$, $M_f=4$;
  random 7-regular and Barab\'asi--Albert networks with $m=4$)}
  \label{fig:si-misinfo-spread}
\end{figure*}

\paragraph{Spread of false information.} At the end of an attack, false information reaches $30$--$66\%$ of the sources' direct neighbours, at most $17\%$ of individuals two steps away, and almost no one further, and it changes the contribution only of FH, since buyers hold the true threshold and UD, UC, and BD ignore messages (Figure~\ref{fig:si-misinfo-spread}). Its reach falls with the share of buyers (Spearman $\rho=-0.77$ on random regular and $-0.71$ on Barab\'asi--Albert networks): disinformation reaches $35\%$ of individuals when buyers are fewer than $5\%$, but $6\%$ when they exceed $15\%$. Relay noise gives many individuals a wrong threshold, but its errors go in both directions, so success does not fall.

\paragraph{Why heterogeneity lowers success here.} In \citet{SantosPacheco2011}, heterogeneous networks favour cooperation in the CRD because fitness accumulates over all games, so well-connected cooperators earn more and are imitated. We average payoffs over groups, which removes this advantage.

\section{Limitations of the network model}
\label{sec:si-limitations}

Attackers and sources are fixed nodes that do not contribute, and receivers do not learn to distrust messages. All networks have $Z=100$ nodes, matching the well-mixed analysis, but at this size, heterogeneous networks cannot display a clear power-law degree distribution. We also do not consider other-regarding preferences and information signals, such as reputation or trust. We leave addressing these shortcomings for future work.

\section{Additional figures}

\begin{figure}[!htbp]
  \centering
  \includegraphics[width=0.85\linewidth]{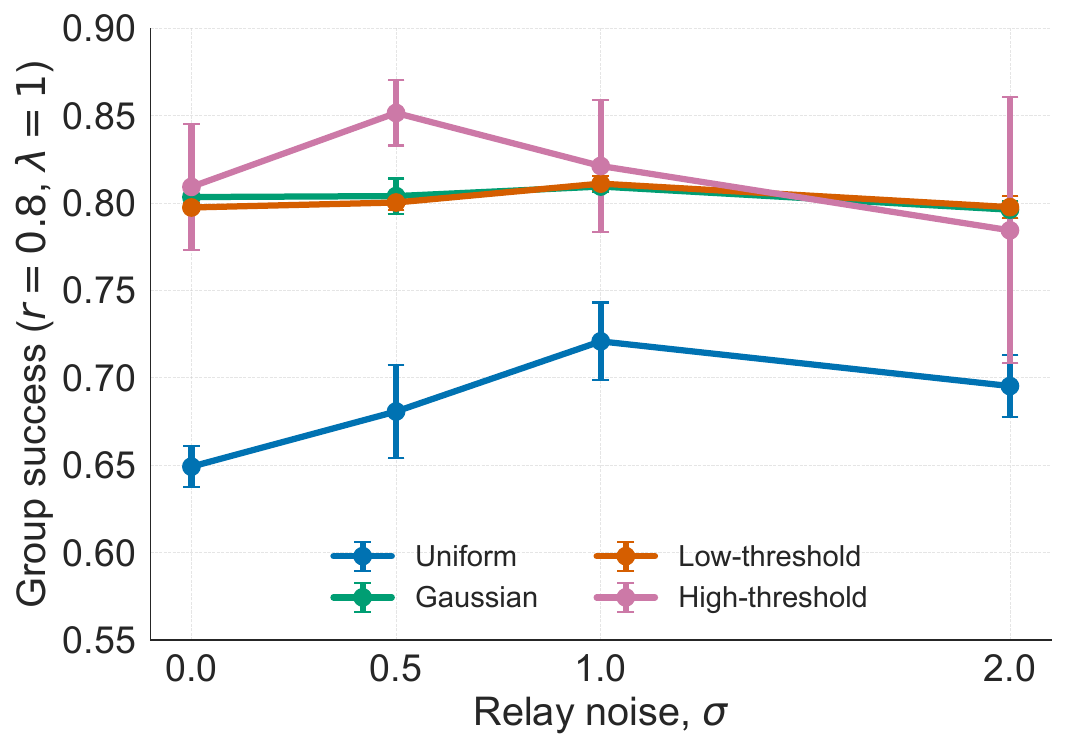}
  \caption{Group success on the random 7-regular network as a function of
  relay noise $\sigma$ ($r=0.8$, $\lambda=1$). Error bars are 95\% intervals
  across chains (10 at $\sigma=0$, on 5 networks, with $3\times10^{4}$
  generations; 6 otherwise, on 3 networks, with $2\times10^{4}$
  generations; 2 initial states per network, after $5\times10^{3}$
  generations of burn-in). ($Z=100$, $N=8$, $\beta=3$,
  $\mu=0.01$, $c=0.1$, $c_I=0.01$, $b=1$, $M_f=4$)}
  \label{fig:si-network-noise}
\end{figure}

\begin{figure}[!htbp]
  \centering
  \includegraphics[width=0.85\linewidth]{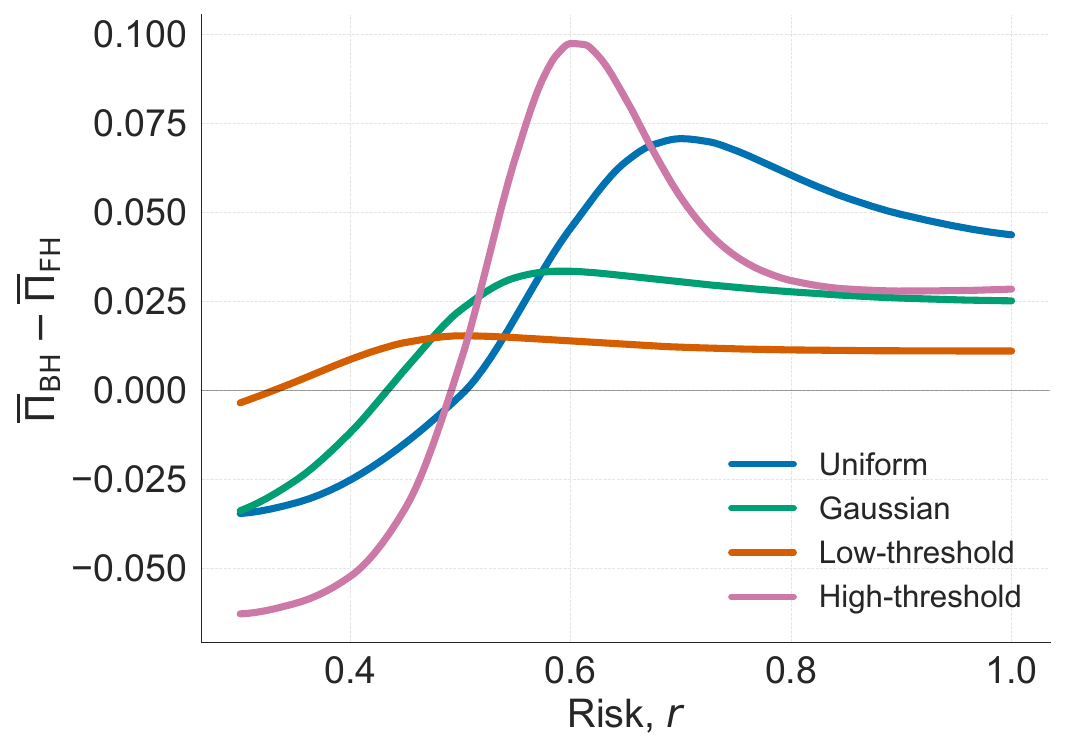}
  \caption{Average payoff of buyers (BH) minus that of information
  free-riders (FH) with sharing. ($Z=100$, $\beta=3$, $\mu=0.01$, $N=8$,
  $c=0.1$, $c_I=0.01$, $b=1$, $M_f=4$)}
  \label{fig:si-shared-info-payoff-gap}
\end{figure}

\begin{figure*}[t]
  \centering
  \includegraphics[width=0.75\textwidth]{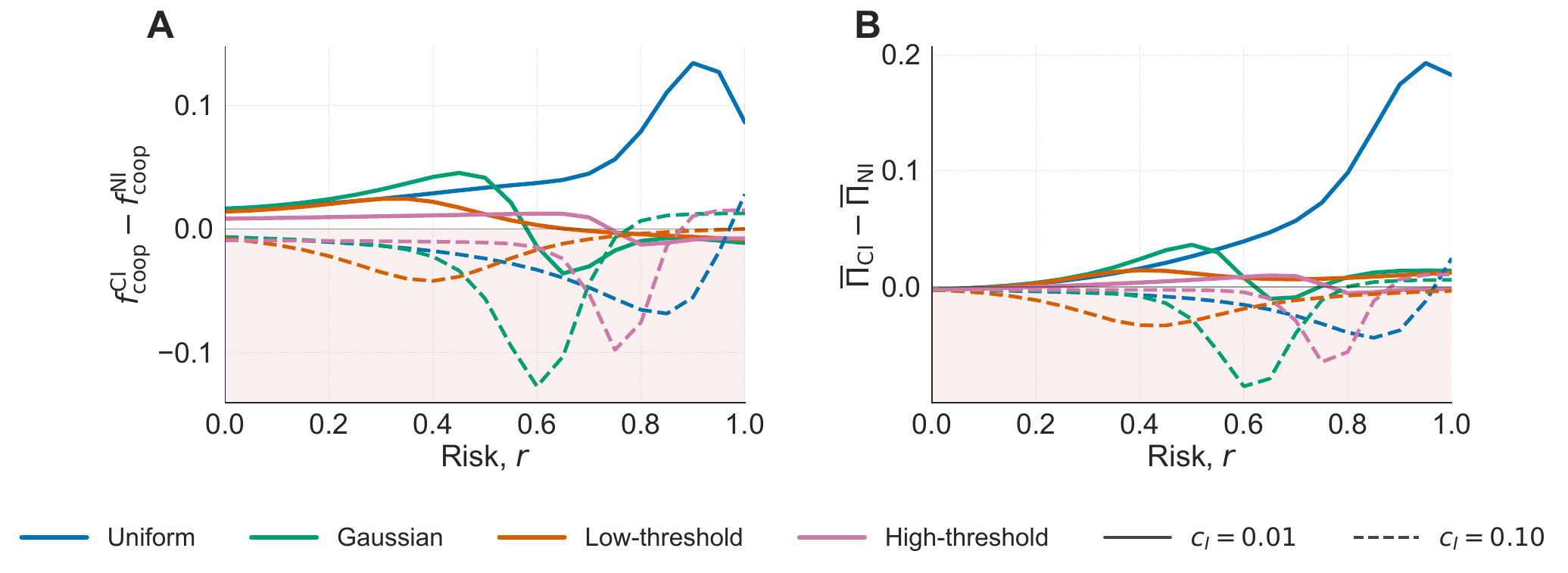}
  \caption{Companion to Figure~\ref{fig:baseline-comparison}: change in
  \textbf{(A)} cooperation frequency and \textbf{(B)} population-average
  payoff relative to the NI baseline, at $c_I=0.01$ (solid) and $c_I=0.10$
  (dashed). Red shading marks where CI performs worse than NI. ($Z=100$,
  $\beta=3$, $\mu=0.01$, $N=8$, $c=0.1$, $b=1$, $M_f=4$)}
  \label{fig:si-coop-payoff-difference}
\end{figure*}

\end{document}